\documentclass[12pt]{article}
\usepackage[top=1in, bottom=1in, left=1.25in, right=1.25in]{geometry}
\usepackage{float}
\usepackage{color}
\usepackage{tikz}
\usepackage{setspace}
\usepackage[toc,page]{appendix}
\usepackage{amssymb}
\usepackage{showlabels}
\usepackage[normalem]{ulem}
\usepackage{multirow}
\usepackage{tikz}
\usepackage[toc]{appendix}
\usepackage{chngcntr}
\usetikzlibrary{shapes,arrows}
\usepackage{subfigure}
\usepackage[round]{natbib}
\usepackage{latexsym}
\usepackage{amsmath}
\usepackage{graphicx}
\usepackage{caption}

\usepackage{bm}
\usepackage[hypertexnames=false,linktocpage=true]{hyperref}
\hypersetup{colorlinks=true,linkcolor=blue,anchorcolor=blue,citecolor=blue,filecolor=blue,urlcolor=blue,bookmarksnumbered=true,pdfview=FitB}

\newcommand{\biblist}{\begin{list}{}
		{\listparindent 0.0cm \leftmargin 0.50cm \itemindent -0.50 cm
			\labelwidth 0 cm \labelsep 0.50 cm
			\usecounter{list}}\clubpanelty4000\widowpanelty4000}
	\newcommand{\ebiblist}{\end{list}}

\newtheorem{theorem}{Theorem}[section]
\newtheorem{lemma}[theorem]{Lemma}

\newtheorem{remark}{Remark}

\def\T{{ \mathrm{\scriptscriptstyle T} }}
\title{\LARGE \bf Semiparametric fractional imputation using Gaussian mixture models for handling multivariate missing data}
\author{Hejian Sang\thanks{ Acknowledgment
		\textit{This work is done when the author was in Iowa State University}}\hspace{.2cm}\\
	Google Inc\\
	and \\
	Jae Kwang Kim\\
	Department of Statistics, Iowa State University,Ames}

\begin{document}
\baselineskip .3in
\maketitle

\begin{abstract}
	Item nonresponse is frequently encountered in practice. Ignoring missing
	data can lose efficiency and lead to misleading inference. Fractional imputation is a frequentist approach of imputation for handling missing data. However, the parametric fractional imputation of \cite{kim2011parametric} may be subject to bias under model misspecification. In this paper, we propose a novel semiparametric fractional imputation method using Gaussian
	mixture models. The proposed method is computationally efficient and leads to robust estimation. The proposed method is further extended to incorporate the categorical auxiliary information. The asymptotic model consistency and $\sqrt{n}$- consistency of the semiparametric fractional imputation estimator are also established. Some simulation studies are presented to check the finite sample performance of the proposed method.
\end{abstract}

\noindent%
\vfill

\newpage

\section{Introduction}\label{sec:intro}

Missing data is frequently encountered in survey sampling, clinical trials and many other areas. 
Imputation can be used to handle item nonresponse and several imputation methods have been developed in the literature.  Motivated from a Bayesian perspective, 
\cite{rubin1996multiple} proposed multiple imputation
to create multiple complete data sets. Alternatively, under the frequentist framework, fractional imputation  \citep{kim2004fractional,kim2011parametric} makes one complete data with multiple imputed values and  their corresponding fractional weights. \cite{little2014statistical} and \cite{kim2013statistical} provided comprehensive overviews of the methods for handling missing data.

For multivariate missing data with arbitrary missing patterns, valid imputation methods should preserve the correlation structure in the imputed data. \cite{judkins2007preservation} proposed an iterative hot deck imputation procedure that is closely related to the data augmentation algorithm of \cite{tanner1987calculation} but they did not provide variance estimation. Other imputation procedures for multivariate missing data include the multiple imputation approaches of \cite{raghunathan2001multivariate} and \cite{,murray2016multiple}, and parametric fractional imputation of \cite{kim2011parametric}. The approaches of \cite{judkins2007preservation} and \cite{raghunathan2001multivariate} are based on conditionally specified models and the imputation from the conditionally specified model is generically subject to the model compatibility problem \citep{chen2010compatibility,liu2013stationary, bartlett2015multiple}. Conditional models for the different missing patterns calculated directly from the observed patterns may not be compatible with each other. The parametric fractional imputation uses the joint distribution to create imputed values and does not suffer from model compatibility problems. 

Note that parametric imputation requires correct model specification. Nonparametric imputation methods, such as kernel regression imputation \citep{cheng1994nonparametric,wang2009empirical},  are robust but may be subject to the curse of dimensionality. Hence, it is important, often critical, to develop a unified, robust and efficient imputation method that can be used for general purpose estimation.
The proposed semiparametric method fills in this important gap by considering a more flexible model for imputation. 

In this paper, to achieve robustness against model misspecification, we develop an imputation procedure based on Gaussian mixture models.  Gaussian mixture model is a very flexible model that can be used to handle outliers, heterogeneity and skewness.   \cite{mclachlan2004finite} and \cite{bacharoglou2010approximation} argued that any continuous distribution can be approximated by a finite Gaussian mixture distribution.  The proposed method using Gaussian mixture model makes a nice compromise between efficiency and robustness. It is semiparametric in the sense that the number of mixture components is chosen automatically from the data. The computation for parameter estimation in our proposed method is based on EM algorithm and its implementation is  relatively simple and efficient.

We note that \cite{di2007imputation} also proposed to use Gaussian mixture model to impute missing data. However, variance estimation and choice of the mixture component are not discussed in \cite{di2007imputation}. 
\cite{elliott2007using} and \cite{kim2014multiple} introduced the multiple imputation using mixture models. Instead of multiple imputation, we use fractional imputation for general-purpose estimation. We provide a completely theoretical justification for consistency of the proposed imputation method. The variance estimator and the model selection for the number of mixture component are also carefully discussed and demonstrated in numerical studies. The proposed method is further extended to handle mixed type data including categorical variable in Section \ref{sec:extension}. By allowing the proportion vector of mixture component to depend on categorical auxiliary variable, the proposed fractional imputation using Gaussian mixture models can incorporate the observed categorical variables and provide a very flexible tool for imputation.

The paper is structured as follows. The setup of the problem is introduced and a short review of fractional imputation are presented in Secction \ref{sec:setup}.  In Section \ref{sec:prop}, the proposed semiparametric method and its algorithm for implementation are introduced. Some asymptotic results are presented in Section \ref{sec:theory}. In Section \ref{sec:extension}, the  proposed method is further extended to handle mixed type data. Some numerical studies and a real data application are presented to show the performance of the proposed method in Section \ref{sec:sim} and Section \ref{sec:app}, respectively. In Section \ref{sec:discuss}, some concluding remarks are made. The technical derivations and proof are presented in Appendix. 


\section{Basic Setup}\label{sec:setup}
Consider a $p$-dimensional vector of study variable $ Y=(y_1, y_2, \cdots, y_p)$. Suppose that $\left\lbrace Y_1, Y_2, \cdots, Y_n\right\rbrace $ are $n$ independent and identically distributed realizations of the random vector $Y$. In this paper, we use the upper case to represent vector or matrix and the lower case to denote the elements within vector or matrix.   Assume that we are interested in estimating parameter $ \theta\in  \Theta$, which is defined through $E\left\lbrace U( \theta; Y)\right\rbrace = 0$, where $U(\cdot; Y)$ is the estimating function of $ \theta$. With no missingness, a consistent estimator of $\theta$ can be obtained by the solution to
\begin{eqnarray}
\frac{1}{n}\sum_{i=1}^{n} U( \theta; Y_i)=0.\label{CC}
\end{eqnarray}
To avoid unnecessary details, we assume that the solution to (\ref{CC}) exists uniquely almost everywhere.

However, due to missingness, the estimating equation in (\ref{CC}) cannot be applied directly. To formulate the multivariate missingness problem, we further define the response indicator vector $R=(r_1, r_2, \cdots, r_p)$ for $Y=(y_1,y_2,\cdots, y_p)$ as 
\begin{eqnarray}
r_j=\left\lbrace 
\begin{array}{ll}
1 & \text{if $y_j$ is observed}\\
0 & \text{otherwise},
\end{array}\right.\nonumber
\end{eqnarray}
where $j=1,2,\cdots, p$. We assume that the response mechanism is missing at random in the sense of \cite{rubin1976inference}. We decompose $ Y=( Y_{obs}, Y_{mis})$, where $Y_{obs}$ and $Y_{mis}$ represent the observed and missing parts of  $Y$, respectively. Thus, the missing-at-random assumption is described as
\begin{eqnarray}
\mathrm{pr}\left\lbrace R=(r_1,r_2, \cdots, r_p)\mid  Y_{obs},  Y_{mis}\right\rbrace =\mathrm{pr}\left\lbrace R=(r_1,r_2, \cdots, r_p)\mid Y_{obs}\right\rbrace , \label{mar}
\end{eqnarray} 
for any $ r_j\in \{0,1\}$, $j=1,2,\cdots, p$.

Under the missing-at-random assumption, a consistent estimator of $\theta$ can be obtained by solving the following estimating equation:
\begin{eqnarray}
\frac{1}{n}\sum_{i=1}^{n}E\left\lbrace U(\theta; Y_i)\mid Y_{i, obs} \right\rbrace=0, \label{eq4}
\end{eqnarray}
where it is understood that $E\left\lbrace U(\theta; Y_i)\mid Y_{i, obs} \right\rbrace=U(\theta; Y_i)$ if $Y_{i, obs}=Y_i$.
To compute the conditional expectation in (\ref{eq4}), the parametric fractional imputation (PFI) method of \cite{kim2011parametric} can be developed. To apply the parametric fractional imputation, we can assume that the random vector $Y$ follows a parametric model in that $F_0(Y)\in \left\lbrace F_{ \zeta}( Y): \zeta \in \Omega\right\rbrace $.  In the parametric fractional imputation, $M$ imputed values for $ Y_{i, mis}$, say $\{ Y_{i, mis}^{*(1)}, Y_{i, mis}^{*(2)}, \cdots, Y_{i, mis}^{*(M)}\} $ are generated from a proposal distribution with the same support of $F_0(Y)$  and are assigned with fractional weights, say $\left\lbrace w_{i1}^{*}, w_{i2}^*, \cdots, w_{iM}^*\right\rbrace $, so that a consistent estimator of $\theta$ can be obtained by solving $$\sum_{i=1}^{n}\sum_{j=1}^{M}w_{ij}^*U( \theta; Y_{i,obs}, Y_{i, mis}^{*(j)})=0,$$
where the fractional weights are constructed to satisfy $\sum_{j=1}^{M}w_{ij}^*U(\theta; Y_{i, obs}, Y_{i, mis}^{*(j)} )\cong E\left\lbrace U\left(\theta; Y_i\right)\mid Y_{i, obs} \right\rbrace$
as closely as possible, with $\sum_{j=1}^M w_{ij}^*=1$. In \cite{kim2011parametric}, importance sampling idea is used to compute the fractional weights.

However, for multivariate missing data, it is not easy to find a joint distribution family $\left\lbrace F_{\zeta}(Y):  \zeta \in \Omega\right\rbrace$ correctly.  If the joint distribution family $\left\lbrace F_{\zeta}( Y): \zeta \in \Omega\right\rbrace$ is misspecified, the parametric fractional imputation can lead to biased inference.  All aforementioned concerns motivate us to consider a more robust fractional imputation method using Gaussian mixture models, which cover a wider class of parametric models.

\section{Proposed method}\label{sec:prop}
We assume that the random vector $Y$ follows a Gaussian mixture model
\begin{eqnarray}
f( Y; \alpha, \zeta)=\sum_{g=1}^{G}\alpha_gf( Y; \zeta_g),\label{mm}
\end{eqnarray}
where $G$ is the number of mixture component, $\alpha_g\in (0,1)$ is the mixture proportion satisfying $\sum_{g=1}^G\alpha_g=1$,  and $f(\cdot; \zeta_g)$ is the density function of multivariate normal distribution with parameter $\zeta_g=\left\lbrace  \mu_g, \Sigma_g\right\rbrace $. Here, we consider  the same values of $\Sigma_g=\Sigma$ across all components to get a parsimonious model. The proposed joint model in (\ref{mm}) can be easily extended to use the group-dependent variance $\Sigma_g$, as in \cite{di2007imputation}. 

To formulate the proposal, define the group indicator vector $ Z=(z_1,z_2,\cdots,z_G)$, where $z_g=1$ and $z_j=0$ for all $j\neq g$, if sample unit belongs to the $g$-th group. Note that $Z$ is a latent variable with parameter $\mathrm{pr}(z_g=1)=\alpha_g$, satisfying $\sum_{g=1}^G\alpha_g=1$.  
We assume that the Gaussian mixture model in (\ref{mm}) satisfies the strong first-order identifiability assumption \citep{chen1995optimal,liu2003asymptotics,chen2008order}, where the first-order derivatives of $f(Y; \alpha, \zeta)$ respect to all parameters are linearly independent.
Using $Z$ variable, we can express $f(Y)=\sum_{g=1}^{G}\mathrm{pr}(z_g=1)f(Y\mid z_g=1)$,
which leads to the marginal distribution in (\ref{mm}).

To  handle item nonresponse, we propose to use the fractional imputation method to impute the missing values. Note that, the joint predictive distribution of $(Y_{mis}, Z)$ given $Y_{obs}$ can be written as $f(Y_{mis},Z\mid Y_{obs})=f(Z\mid Y_{obs})f(Y_{mis}\mid Y_{obs}, Z)$,
which implies that the prediction model for $Y_{mis}$ is 
\begin{eqnarray}
f(Y_{mis}\mid Y_{obs})=\sum_{g=1}^G\mathrm{pr}(z_{g}=1\mid Y_{obs})f(Y_{mis}\mid Y_{obs}, z_{g}=1).\label{predictive}
\end{eqnarray}
The first part in (\ref{predictive}) can be obtained by 
\begin{eqnarray}
\mathrm{pr}(z_{g}=1\mid Y_{obs})=\frac{f(Y_{obs}\mid z_{g}=1 )\alpha_g}{\sum_{g=1}^{G}f(Y_{obs}\mid z_{g}=1 )\alpha_g}\label{p(z)},\nonumber
\end{eqnarray}
where $Y_{obs}\mid( z_{g}=1)$ is normal. The second part of (\ref{predictive}) , which is $Y_{mis}\mid (Y_{obs}, z_{g}=1)$, is also normal. 
Therefore, the EM algorithm for the proposed fractional imputation using Gaussian mixture models (FIGMM) can be described as follows:
\begin{itemize}
	\item []\textit{I-step}: To generate $Y_{i,mis}^*$ from $f(Y_{i,mis}\mid Y_{i, obs}; \alpha^{(t)}, \zeta^{(t)})$ in (\ref{predictive}), we use the following two-step method:
	\begin{itemize}
		\item []\textit{Step 1}: Compute 
		\begin{eqnarray}
		p_{ig}^{(t)}= \frac{f(Y_{i,obs}\mid z_{ig}=1; \zeta_g^{(t)})\alpha_g^{(t)}}{\sum_{g=1}^G f(Y_{i,obs}\mid z_{ig}=1; \zeta_g^{(t)})\alpha_g^{(t)}},\nonumber
		\end{eqnarray}
		where $f(Y_{i, obs}\mid z_{ig}=1; \zeta_g)$ is the marginal density of $Y_{i, obs}$ derived from $(Y_{i, obs}, Y_{i, mis})\mid (z_{ig}=1)\sim N(\mu_g, \Sigma)$.  Generate $(M_{i1}^{(t)}, M_{i2}^{(t)}, \cdots, M_{iG}^{(t)})\sim \mathrm{Multinomial}(M; { \mathrm P_i}^{(t)})$, where ${ \mathrm P_i}^{(t)}=(p_{i1}^{(t)}, \cdots, p_{iG}^{(t)})$
		with $M_{ij}^{(t)}>0$.
		\item[] \textit{Step 2}:
		For each $g=1,2,\cdots, G$, we generate $M_{ig}^{(t)}$ independent realizations of $Y_{i, mis}^*$, say $\{Y_{i, mis}^{*(gj)}, j=1,2,\cdots, M_{ig}^{(t)}\}$, from the conditional distribution $f(Y_{i,mis}\mid Y_{i, obs},z_{ig}=1; \zeta_{g}^{(t)})$, which is also normal.
		
	\end{itemize}
	\item[] \textit{W-step}: Compute the fractional weights for $Y_{i,mis}^{*(gj)}$ as $w_{igj(t)}^*=p_{ig}^{(t)}/M_{ig}^{(t)}$.
	Note that $\sum_{g=1}^G\sum_{j=1}^{M_{ig}^{(t)}} w_{igj(t)}^*=1$, for each $i=1,2,\cdots,n$.
	Using $(w_{ig j(t)}^*, Y_{i,mis}^{*(gj)})$, we can compute
	\begin{eqnarray}
	Q^*(\alpha, \zeta\mid \alpha^{(t)}, \zeta^{(t)})=\sum_{i=1}^{n}\sum_{g=1}^{G}\sum_{j=1}^{M_{ig}^{(t)}} w_{igj(t)}^*
	\left\lbrace \log \alpha_g +\log f(Y_{i}^{*(gj)}\mid z_{ig}=1; \zeta_g)\right\rbrace, \label{Q_star}
	\end{eqnarray}
	where $Y_{i}^{*(gj)}=(Y_{i, obs}, Y_{i, mis}^{*(gj)})$. If $\delta_i=1$, then $Y_{i}^{*(gj)}=Y_i$.
	\item[]\textit{M-step}: Update the parameters by maximizing (\ref{Q_star}) with respect to $(\alpha, \zeta)$. The solution is 
	\begin{eqnarray}
	\alpha_{g}^{(t+1)}=\frac{1}{n}\sum_{i=1}^{n}\frac{M_{ig}^{(t)}}{M};\qquad 
	\mu_g^{(t+1)}=\frac{\sum_{i=1}^n \sum_{j=1}^{M_{ig}^{(t)}}w_{igj(t)}^* Y_{i}^{*(gj)}}{\sum_{i=1}^n \sum_{j=1}^{M_{ig}^{(t)}} w_{igj(t)}^*};\nonumber\\
	\Sigma^{(t+1)}=\frac{1}{n}\sum_{i=1}^n \sum_{g=1}^G\sum_{j=1}^{M_{ig}^{(t)}} w_{igj(t)}^*\left( Y_{i}^{*(gj)}-\mu_g^{(t+1)}\right)\left( Y_{i}^{*(gj)}-\mu_g^{(t+1)}\right)^\T.\nonumber
	\end{eqnarray}
\end{itemize}
Repeat \textit{I-step} to \textit{M-step} until the convergence is achieved.
Then, the final estimator, say $\hat{ \theta}_{FIGMM}$, of $ \theta$ can be obtained by solving the fractionally imputed estimating equation, given by
\begin{eqnarray}
\frac{1}{n}\sum_{i=1}^{n}\sum_{g=1}^G\sum_{j=1}^{M_{ig}}w_{igj}^*U(\theta; Y_{i}^{*(gj)}) =0,\label{est}
\end{eqnarray}
where $w_{igj}^*$ are the final fractional weights and $M_{ig}$ are the final imputation sizes for group $g$, after convergence of the EM algorithm.

\begin{remark}\label{rk2}
	We now briefly discuss variance estimation of $\hat\theta_{FIGMM}$. To estimate the variance of $\hat\theta_{FIGMM}$, replication methods, such as jackknife, can be used. First note that, the fractional weight assigned to $Y_{i}^{*(gj)}$ is $	w_{igj}^*=\hat p_{ig}M_{ig}^{-1}:=\hat p_{ig}\hat \pi_{2j\mid ig}$,
	where $\hat p_{ig}$ is obtained from 
	\begin{eqnarray}
	\hat p_{ig}=\frac{f(Y_{i, obs}\mid z_{ig}=1; \hat \zeta_g)\hat \alpha_g}{\sum_{g=1}^G f(Y_{i, obs}\mid z_{ig}=1; \hat \zeta_g)\hat \alpha_g}.\label{weightp}
	\end{eqnarray}
	Thus, the $k$-th replicate of $w_{igj}^*$ can be obtained by 
	\begin{eqnarray}
	w_{igj}^{*(k)}=\hat p_{ig}^{(k)}\hat \pi_{2j\mid ig}^{(k)},\label{wigjtk}
	\end{eqnarray}
	where $\hat p_{ig}^{(k)}$ is obtained from (\ref{weightp}) using $\hat \zeta^{(k)}$ and $\alpha_g^{(k)}$, the $k$-th replicate of $\hat\zeta$ and $\hat \alpha_g$ respectively, and 
	\begin{eqnarray}
	\hat \pi_{2j\mid ig}^{(k)}\propto \frac{f(Y_{i, mis}^{*(gj)}\mid Y_{i, obs}, z_{ig}=1; \hat \zeta_g^{(k)})}{f(Y_{i, mis}^{*(gj)}\mid Y_{i, obs}, z_{ig}=1; \hat\zeta_g)}\nonumber
	\end{eqnarray}
	and $\sum_{j=1}^{M_{ig}} \hat \pi_{2j\mid ig}^{(k)}=1$. The calculation of $\hat \pi_{2j\mid ig}^{(k)}$ is based on the idea of importance sampling. Construction of replicate fractional weights using importance sampling idea has been used in \cite{berg2016imputation}. 
	
	The replicate parameter estimates $(\hat \alpha^{(k)},\hat \zeta^{(k)})$ are computed by maximizing
	\begin{eqnarray}
	l_{obs}^{(k)}(\alpha, \zeta)=\sum_{i=1}^{n}w_{i}^{(k)} \log f_{obs}(Y_{i, obs}; \alpha, \zeta) \label{lobs}
	\end{eqnarray}
	respect to $(\alpha, \zeta)$, where $	f_{obs}(Y_{i, obs}; \alpha, \zeta)=\sum_{g=1}^G \alpha_g f(Y_{i, obs}\mid z_{ig}=1; \zeta_g)$,
	and $w_{i}^{(k)}$ is the $k$-th replicate of $w_i=n^{-1}$. The maximizer of $l_{obs}^{(k)}(\alpha, \zeta)$ in (\ref{lobs}) can be obtained by applying the same EM algorithm using replicate weights and replicate fractional weights in the \textit{W-step} and \textit{M-step}. There is no need to repeat \textit{I-step}. Variance estimation for $\hat \theta_{FIGMM}$ can be obtained by computing the $k$-th replicate of $\hat\theta_{FIGMM}$ from 
	\begin{eqnarray}
	\sum_{i=1}^{n}w_{i}^{(k)}\sum_{g=1}^{G}\sum_{j=1}^{M_{ig}}w_{igj }^{*(k)} U(\theta;Y_{i}^{*(gj)})=0. \label{jk}
	\end{eqnarray}
	For example, the jackknife variance estimator of $\hat \theta_{FIGMM}$ can be obtained by 
	\begin{eqnarray}
	\widehat {\mathrm{var}}\left( \hat \theta_{FIGMM}\right) =\frac{n-1}{n}\sum_{i=1}^n\left(\hat \theta_{FIGMM}^{(k)}-\hat \theta_{FIGMM} \right)^2,\nonumber
	\end{eqnarray}
	where $\hat \theta_{FIGMM}^{(k)}$ is the $k$-th replicate of $\hat\theta_{FIGMM}$ obtained from (\ref{jk}).
\end{remark}


\section{Asymptotic theory}\label{sec:theory}
In our proposed fractional imputation method using Gaussian mixture models in \S \ref{sec:prop}, we have assumed that the size of mixture components, $G$, is known. In practice , $G$ is often unknown and we need to estimate it from the sample data. If $G$ is larger than necessary, the proposed mixture model may be subject to overfitting and increase its variance. If $G$ is small, then the approximation of the true distribution cannot provide accurate prediction due to its bias. Hence, we can allow the model complexity parameter $G$ to depend on the sample size $n$, say $G=G(n)$. The choice of $G$ under complete data has been well explored in the literature. The  popular methods are based on Bayesian information criterion (BIC) and Akaike's information criterion (AIC). See \cite{ wallace1999minimum}, \cite{windham1992information}, \cite{schwarz1978estimating}, \cite{fraley1998many}, \cite{keribin2000consistent} and \cite{dasgupta1998detecting}. The alternative way of using SCAD penalty \citep{fan2001variable} is studied in \cite{chen2008order} and \cite{huang2017model}.

In this paper,  we consider using the Bayesian information criterion to select $G$. Under multivariate missingness, we do not have the complete log-likelihood function. Thus, we  use the observed log-likelihood function to serve the role of the complete log-likelihood function in computing the information criterion, in the sense that 
\begin{eqnarray}
\mathrm{BIC}(G)=-2\sum_{i=1}^{n}\log \left\lbrace \sum_{g=1}^{G}\hat \alpha_g f(Y_{i,obs}\mid z_{ig}=1;\hat \zeta_g)\right\rbrace + (\log n)\phi(G),\label{BIC2}
\end{eqnarray}
under the assumption of $\Sigma_g=\Sigma$, where $(\hat \alpha, \hat \zeta)$ are the estimators obtained from the proposed method and $\phi(G)$ is a monotone increasing function of $G$. In (\ref{BIC2}), $\phi(G)=G+Gp$ if ignoring constant terms. However, our model selection framework and theoretical results can be directly applied to any general penalty function $\phi(G)$.
Using Gaussian mixture models, the observed log-likelihood function is expressed as a closed-form.

In this section, we first establish the consistency of model selection using (\ref{BIC2}) under the Gaussian mixture model assumption.  After that, we establish some asymptotic results when the Gaussian mixture model assumption is violated.

To establish the first part, assume that $\{Y_1, Y_2, \cdots, Y_n\}$ is a random sample  from 
$f_0(Y)=\sum_{g=1}^{G^o}\alpha_{g}^of(Y; \zeta_{g}^o)$, where $(G^o, \alpha^o, \zeta^o)$ are true parameter values. For $\zeta_g^o=(\mu_g^o, \Sigma^o)$, we need the following regularity assumptions:
\begin{itemize}
	\item[] (\textit{A1}) The mean vectors for each mixture component is bounded uniformly, in the sense of $\|\mu_g^o\|\leq C_1$, for $g=1,2,\cdots, G^o$.\label{A1}
	\item[](\textit{A2}) $\|\Sigma^o\|\leq  C_2$. Furthermore, $\Sigma^o$ is nonsingular.\label{A2}
\end{itemize}

The first assumption means the first moment is bounded. Assumption (\textit{A2}) is to make sure that $\Sigma^0$ is bounded and nonsingular. Both assumptions are commonly used. 

To establish the model consistency, we furthermore make the additional assumptions on the response mechanism: 
\begin{itemize}
	\item [](\textit{A3}) The response rate for $y_j$ is bounded below from 0, say $\lim_{n}n^{-1}\sum_{i=1}^n r_{ij}>C_3$ , for $j=1,2,\cdots, p$, where $C_3>0$ is a constant.\label{A3}
	\item [](\textit{A4}) The response mechanism satisfies the mising-at-random condition in (\ref{mar}).\label{A4}
\end{itemize}

The following theorem shows that the true number of mixture components can be selected by minimizing $\text{BIC}(G)$ in (\ref{BIC2}) consistently. 

\begin{theorem}\label{thm1}
	Assume the true density $f_0$ is the Gaussian mixture model, satisfying (\textit{A1})--(\textit{A2}).
	Let $\hat G$ be the minimizer of $\mathrm{BIC}(G)$ in (\ref{BIC2}).
	Under assumptions (\textit{A3})--(\textit{A4}), we have 
	\begin{eqnarray}
	\mathrm{pr}(\hat G=G^o)\xrightarrow{}1,\nonumber
	\end{eqnarray}
	as $n\xrightarrow{}\infty$, where $G^o$ is the true number of mixture components.
\end{theorem}
The proof of Theorem \ref{thm1} is shown in the Supplementary Material. Theorem \ref{thm1} states that minimizing $\text{BIC}(G)$ consistently selects the true mixture components under the assumption that the true distribution is in the Gaussian mixture model. 

Now, in the second scenario, the true distribution does not necessary belong to the class of Gaussian mixture models, Thus, we first establish the following lemma to measure how well Gaussian mixture model can approximate the arbitrary density function.  We furthermore make additional assumptions about the true density function $f_0$.  Use $E_0$ to denote the expectation respect to $f_0$.

\begin{itemize}
	\item [](\textit{A5}) Assume $f_0(Y)$ is continuous with $\int Y^2 f_0(Y)<\infty$.
	\item[](\textit{A6}) Assume $E_0\left\lbrace \partial f(Y)/\partial\alpha\right\rbrace <\infty$ and $E_0\left\lbrace \partial f(Y)/\partial\mu\right\rbrace <\infty$, where $f(Y)=\sum_{g=1}^{G}\alpha_gf (Y;\mu_g, \Sigma)$. Moreover, assume $E_0\left\lbrace f(Y)^{-2}\right\rbrace<\infty $.
\end{itemize}

Assumption (\textit{A5}) is satisfied for any continuous random variable with bounded second moments. Assumption (\textit{A6}) is true for any finite Gaussian mixture model and $f_0$ has a valid moment generating function.
\begin{lemma}\label{lemma_f}
	Under assumptions (\textit{A5})--(\textit{A6}) and missing at random, for any $\epsilon>0$, there exist $G=\epsilon^{-\gamma}$, such that 
	\begin{eqnarray}
	\|f_0-\hat f\|_1=O(\epsilon)\label{L1},\\
	\mathrm{var}(f_0-\hat f)=O(\epsilon^{-\gamma}n^{-1}),\label{L2}
	\end{eqnarray}
	with probability one, where $\hat f(Y)=\sum_{g=1}^{G}\hat \alpha_gf (Y;\hat \mu_g, \hat \Sigma)$ is obtained from the proposed method in \S \ref{sec:prop}, $\gamma> 0$ and $\|f_0-\hat f\|_1=\int |f_0(Y)-\hat f(Y)|f_0(Y) \mathrm{d}Y.$
\end{lemma}
The proof of Lemma \ref{lemma_f} is presented in the Supplementary Material.
If $f_0$ is a density function of the Gaussian mixture model, then $\gamma=0$ and by Theorem \ref{thm1}, our proposed $\mathrm{BIC}(G)$ can select the true model consistently.  For any $f_0$ satisfies (\textit{A5})--(\textit{A6}) and is not a finite Gaussian mixture model, the bias can goes to 0 as $G\xrightarrow{}\infty$ from (\ref{L1}). The variance will increase as $G\xrightarrow{}\infty$ from (\ref{L2}) for fixed $n$. There is a trade-off between bias and variance for the divergence case $(\gamma>0, G\xrightarrow{}\infty)$.

Using Lemma \ref{lemma_f}, we can further establish the $\sqrt{n}$-consistency of $\hat\theta_{FIGMM}$. The following assumptions are the sufficient conditions to obtain the $\sqrt{n}$-consistency.

\begin{itemize}
	\item [](\textit{A7}) $E_0\left\lbrace U^2(\theta; Y_i)\right\rbrace<\infty$.
	\item [](\textit{A8}) $\gamma \in (0,2)$. 
	\item[] (\textit{A9}) $\epsilon=O(n^{-1/(2-\Delta)})$, for any $\Delta\in (0,2)$.
\end{itemize}
\begin{theorem}\label{Theorem3}
	Under assumptions (\textit{A5})--(\textit{A9}), $\gamma+\Delta<2$ and MAR, we have 
	\begin{eqnarray}
	\frac{1}{n}\sum_{i=1}^{n}\sum_{g=1}^G\sum_{j=1}^{M_{ig}}w_{igj}^*U(\theta; Y_{i}^{*(gj)})\cong J_1+o_p(n^{-1/2}),\label{thm3}
	\end{eqnarray}
	where $J_1=n^{-1}\sum_{i=1}^{n}E_0\left\lbrace U(\theta; Y_i)\mid Y_{i, obs} \right\rbrace$,
	if $M=\sum_{i,g}\{M_{ig}\}\xrightarrow{}\infty$. Furthermore, we have
	\begin{eqnarray}
	\sqrt{n}(\hat \theta_{FIGMM}-\theta_0)\xrightarrow{}N(0, \Sigma),\label{norm}
	\end{eqnarray}
	for some $\Sigma$ which is positive definite and $\theta_0$ satisfies $E_0\left\lbrace U(\theta_0; Y) \right\rbrace=0. $
\end{theorem}
The proof of (\ref{thm3}) is shown in the Supplementary Material and (\ref{norm}) can be directly derived from (\ref{thm3}).
From Theorem \ref{Theorem3}, we have $G=O(n^{\gamma/(2-\Delta)})=o(n)\xrightarrow{}\infty$ with the rate smaller than $n$. Thus, even under non-Gaussian mixture families, our proposed method still enjoys $\sqrt{n}$-consistency. 

\section{Extension}\label{sec:extension}
In Section \ref{sec:prop}, we assume that $Y$ is fully continuous. However, in practice, categorical variables can be used to build imputation models. 
We extend our proposed method to incorporate the categorical variable as a covariate in the model.

To introduce the  proposed  method, we first introduce the conditional Gaussian mixture model.  Suppose that $(X, Y)$ is a random vector where $X$ is discrete and $Y$ is continuous. We further assume that $X$ is always observed.
To obtain the conditional Gaussian mixture model, we  assume that $Z$ satisfies 
\begin{eqnarray}
f( Y\mid X,  Z)=f( Y\mid Z),\label{C1}
\end{eqnarray}
in the sense that $Z$ is a partition of the sample such that $Y$ is homogeneous within each group defined by $Z$. Furthermore, we assume that $f(Y\mid z_g=1)$ follows a Gaussian distribution.  Combining these assumptions, we have the following conditional Gaussian mixture model
\begin{eqnarray}
f(Y\mid X)=\sum_{g=1}^{G}\alpha_g(X)f(Y\mid z_g=1),\label{extend}
\end{eqnarray}
where $\alpha_g(X)=\mathrm{pr}(z_g=1\mid X)$ and $f(Y\mid z_g=1)$ is the density function of the normal distribution with parameter $\zeta_g=\left\lbrace \mu_g, \Sigma\right\rbrace $. We also assume that the identifiability conditions in (\ref{extend}) hold.

Using the argument similar to (\ref{predictive}), the predictive model of $Y_{i, mis}$  under (\ref{C1}) can be expressed as
\begin{eqnarray}
f(Y_{i, mis}\mid Y_{i,obs},X_i)=\sum_{g=1}^{G}\mathrm{pr}(z_{ig}=1\mid Y_{i, obs},X_i)f(Y_{i,mis}\mid Y_{i, obs}, z_{ig}=1),\label{pred2}
\end{eqnarray}
where $f(Y_{i,mis}\mid Y_{i, obs}, z_{ig}=1)$ can be derived from $(Y_{i,obs}, Y_{i, mis})\mid( z_{ig}=1)\sim N(\mu_g, \Sigma)$. The posterior probability of $z_{ig}=1$ given the observed data is
\begin{eqnarray}
\mathrm{pr}(z_{ig}=1\mid Y_{i, obs},X_i)=\frac{f(Y_{i, obs}\mid z_{ig}=1)\mathrm{pr}(z_{ig}=1\mid X_i)}{\sum_{g=1}^G f(Y_{i, obs}\mid z_{ig}=1)\mathrm{pr}(z_{ig}=1\mid X_i)}.\nonumber
\end{eqnarray}

Therefore, the proposed fractional imputation using conditional Gaussian mixture models can be summarized as follows:
\begin{itemize}
	\item []\textit{I-step}: Creating $M$ imputed values of $Y_{i,mis}$ from (\ref{pred2}) can be described in the following two steps.
	\begin{itemize}
		\item []\textit{Step 1}: For each $g=1,2,\cdots, G$, given the current parameter values $(\alpha_{g}^{(t)}, \zeta_g^{(t)})$, the posterior probabilities of $z_{ig}=1$ given $ (Y_{i, obs},X_i)$ can be obtained from 
		\begin{eqnarray}
		p_{ig}^{(t)}=\frac{f(Y_{i, obs}\mid z_{ig}=1; \zeta_g^{(t)})\alpha_{g}^{(t)}(X_i)}{\sum_{g=1}^G f(Y_{i, obs}\mid z_{ig}=1; \zeta_g^{(t)})\alpha_{g}^{(t)}(X_i)}.\nonumber
		\end{eqnarray}
		
		\item []\textit{Step 2}: Generate $M$ imputed values of $Y_{i,mis}$ following the same procedure of \textit{I-step} in \S \ref{sec:prop}. 
	\end{itemize}
	\item[] \textit{W-step}: Update the fractional weights for $Y_{i}^{*(gj)}=(Y_{i, obs}, Y_{i, mis}^{*(gj)})$ as $w_{igj(t)}^*=p_{ig}^{(t)}/{M_{ig}^{(t)}}$,
	for $j=1,2, \cdots M_{ig}^{(t)}$ and $\sum_{g=1}^G M_{ig}^{(t)}=M$. Note that $\sum_{g=1}^G\sum_{j=1}^{M_{ig}^{(t)}} w_{igj(t)}^*=1$.
	
	\item[]\textit{M-step}: Update the parameter values by maximizing
	\begin{eqnarray}
	Q^*(\alpha, \zeta\mid \alpha^{(t)}, \zeta^{(t)})=\sum_{i=1}^{n}\sum_{g=1}^{G}\sum_{j=1}^{M_{ig}^{(t)}} w_{igj(t)}^*
	\left\lbrace \log \alpha_g(X_i) +\log f(Y_{i}^{*(gj)}\mid z_{ig}=1; \zeta_g)\right\rbrace,\nonumber
	\end{eqnarray}
	respect to $(\alpha, \zeta)$.
\end{itemize}

Repeat \textit{I-step} to \textit{M-step} iteratively until convergence is achieved. The final estimator of $\theta$ can be obtained by solving  the fractionally imputed estimating equation in (\ref{est}).
Note that the proposed method builds the proportion vector of mixture components into a function of auxiliary variable and assumes that the mixture components share the same mean and variance structure. Thus, the proposed method can borrow information across different $X$ values. Moreover, the auxiliary information is incorporated to build a more flexible class of joint distributions.

\section{Numerical Studies}\label{sec:sim}

We consider two simulation studies to evaluate the performance of the proposed methods. The first simulation study is used to check the performance of the proposed imputation method using Gaussian mixture models under multivariate continuous variables. The second simulation study considers the case of multivariate mixed categorical and continuous variables. To save space, we only present the first simulation study. The second simulation study is presented in the Supplementary Material.

In the first simulation study, we consider the following models for generating $Y_i=(Y_{i1}, Y_{i2}, Y_{i3})$.
\begin{itemize}
	\item [1.]\textit{M1}:  A mixture distribution with density $f(Y)=\sum_{g=1}^3 \alpha_g f_g(Y)$, where  $(\alpha_1,\alpha_2,\alpha_3)=(0\mbox{$\cdot$}3,0\mbox{$\cdot$}3,0\mbox{$\cdot$}4)$ and $f_g(Y)$ is a density function for multivariate normal distribution with mean $\mu_g$ and variance
	\begin{eqnarray}
	\Sigma(\rho)=\left( \begin{matrix}
	1 & \rho & \rho^2 \\
	\rho & 1 & \rho &\\
	\rho^2 & \rho & 1\\
	\end{matrix}\right).\nonumber
	\end{eqnarray}
	Let $\rho=0\mbox{$\cdot$}7$ and $\mu_1=(-3,-3,-3,), \mu_2=(1,1,1), \mu_3=(5,5,5)$.
	
	\item[2.] \textit{M2}: Use the same  model as \textit{M2} except for $f_1(Y)$, where $f_1(Y)$ is a product of the density for the exponential distribution with rate parameter 1.
	
	\item[3.] \textit{M3}:  $Y_{i1}=1+e_{i1}, Y_{i2}=0\mbox{$\cdot$}5Y_{i1}+e_{i2}$ and  $Y_{i3}=Y_{i2}+e_{i3}$, where $e_{i1}, e_{i2}, e_{i3}$ are independently generated from $N(0,1)$, $\text{Gamma}(1,1)$ and $\chi_1^2$ distributions, respectively.
	\item[4.]\textit{M4}:  Generate $(Y_{i1},Y_{i2})$ independently from a Gaussian distribution with  mean $(1,2)$ and variance 
	\begin{eqnarray}
	\left( \begin{matrix}
	1 & 0\mbox{$\cdot$}5\\
	0\mbox{$\cdot$}5 & 1
	\end{matrix}\right).\nonumber
	\end{eqnarray}
	Let $Y_{i3}=Y_{i2}^2+e_{i3}$, where $e_{i3}\sim N(0,1)$.
\end{itemize}

In \textit{M1}, a Gaussian mixture model with $G=3$ is used to generate the samples.  A non-Gaussian mixture distribution is used in \textit{M2} to check the robustness  of the imputation methods. \textit{M3} and \textit{M4} are used to check the performance of the imputation methods under skewness and nonlinearity, respectively.

The  size for each realized sample is $n= 500$. Once the complete sample is obtained, for $y_{ij}, j=2,3$, we select 25\% of the sample independently to make missingness with the selection probabilities equal to $\pi_{ij}$, where $\mathrm{logit}(\pi_{i2})=-0\mbox{$\cdot$}8+0\mbox{$\cdot$}4y_{i1}$, $	\mathrm{logit}(\pi_{i3})=0\mbox{$\cdot$}4-0\mbox{$\cdot$}8y_{i1}$,
and $\mathrm{logit}(u)=\exp(u)/\left\lbrace 1+\exp(u)\right\rbrace $.
Since we assume $y_{i1}$ are fully observed, the response mechanism is missing at random.

The overall missing rate is approximate 55\%.
For each realized incomplete samples, we apply the following methods:
\begin{itemize}
	\item[][\textit{Full}]: As a benchmark, we use the full samples to estimate parameters. Confidence intervals with 95\% coverage rates are constructed using Wald method.
	\item[] [\textit{CC}]: Only use the complete cases to estimate parameters and construct confidence intervals.
	\item [][\textit{MICE}]: Apply multivariate imputation by chained equations \citep{buuren2011mice}. The predictive mean matching is used as a default. The variance estimators are obtained using Rubin's formula and confidence intervals are constructed using Wald method.
	\item[] [\textit{PFI}]: Parametric fractional imputation method of \cite{kim2011parametric}, where we assume the joint distribution is a multivariate normal distribution.
	\item[] [\textit{SFI}]: The proposed semiparametric fractional imputation method using Gaussian mixture models, where the number of components $G$ is selected using the BIC in (\ref{BIC2}). The confidence interval is implemented using the variance estimator presented in Remark \ref{rk2}.
	
\end{itemize}
The parameters of interest are the population means and the population proportions. For $Y=(y_1,y_2,y_3)$, define $\theta_2=E(y_2), \theta_3=E(y_3)$. The parameters of population proportions are defined as $P_2=\mathrm{pr}(y_2<c_2),P_3=\mathrm{pr}(y_3<c_3)$, where $(c_2,c_3)=(-2,-2)$ for \textit{M1} and \textit{M2}, $(c_2, c_3)=(2,3)$ for \textit{M3}, $(c_2,c_3)=(2,5)$ for \textit{M4}.
The simulation is repeated for $B=2,000$ times.

\begin{figure}
	\centering
	\includegraphics[width=0.4\textwidth]{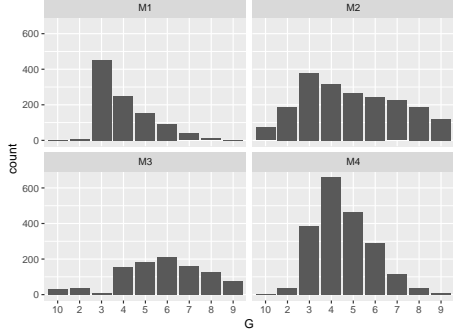}
	\caption{The histograms of selected $G$ values using the proposed BIC.}\label{G2}
\end{figure}

To evaluate the above methods, the relative mean square error (RMSE) is defined as
\begin{eqnarray}
\mathrm{RMSE}=\frac{\mathrm{MSE}_{\text{method}}}{\mathrm{MSE}_{\text{Full}}}\times 100,\label{rmse}\nonumber
\end{eqnarray}
where $\mathrm{MSE}_{\text{method}}$ is the mean squared error of the parameters of the current method and $\mathrm{MSE}_{\text{Full}}$ is the mean squared error of the parameters of using full samples. The simulation results of RMSE and average coverage probability are presented in Table \ref{tbl11}. The histograms of selected $G$ values using the proposed BIC method are shown in Figure \ref{G2}.  

\begin{table}[ht]
	\centering
	\caption{The relative mean squared errors and coverage probabilities (in parentheses) of the five methods computed from a Monte Carlo simulation of size $B=2,000$. Full, the method uses full samples; CC, the method only uses complete cases;  MICE, multiple imputation uses chain equations;  PFI, parametric fractional imputation;  SFI, the proposed semiparametric fractional imputation method uses Gaussian mixture models.}\label{tbl11}
	\begin{tabular}{l|l|rrrr}
		\hline
		Model & Method  & $\theta_2$ & $\theta_3$  & $P_2$ & $P_3$ \\
		\hline
		\multirow{5}{*}{\textit{M1}} & Full & 100(94$\cdot$9) & 100(95$\cdot$0) & 100(94$\cdot$5) & 100(94$\cdot$9) \\ 
		& CC & 3233(2$\cdot$9) & 3010(3$\cdot$5) & 2639(4$\cdot$5) & 2407(5$\cdot$6) \\ 
		& MICE & 100(95$\cdot$2) & 105(93$\cdot$9) & 102(94$\cdot$7) & 145(90$\cdot$6) \\ 
		& PFI & 100(94$\cdot$6) & 104(93$\cdot$3) & 102(94$\cdot$0) & 141(89$\cdot$1) \\ 
		& SFI & 100(94$\cdot$6) & 106(95$\cdot$0) & 102(95$\cdot$3) & 134(95$\cdot$3) \\ 
		\hline
		\multirow{5}{*}{\textit{M2}} & Full & 100(94$\cdot$1) & 100(94$\cdot$6) & 100(94$\cdot$1) & 100(94$\cdot$5) \\ 
		& CC & 2841(3$\cdot$0) & 2863(3$\cdot$6) & 2284(4$\cdot$8) & 2233(5$\cdot$7) \\ 
		& MICE & 102(94$\cdot$4) & 109(93$\cdot$7) & 103(93$\cdot$8) & 189(86$\cdot$3) \\ 
		& PFI & 102(93$\cdot$8) & 108(94$\cdot$1) & 102(93$\cdot$3) & 180(83$\cdot$8) \\ 
		& SFI & 106(95$\cdot$1) & 107(95$\cdot$4) & 103(93$\cdot$8) & 177(92$\cdot$8) \\ 
		\hline
		\multirow{5}{*}{\textit{M3}} & Full & 100(94$\cdot$7) & 100(94$\cdot$7) & 100(95$\cdot$2) & 100(95$\cdot$1) \\ 
		& CC & 157(77$\cdot$9) & 155(85$\cdot$6) & 145(77$\cdot$0) & 145(86$\cdot$6) \\ 
		& MICE & 100(94$\cdot$8) & 100(93$\cdot$5) & 117(92$\cdot$5) & 128(90$\cdot$4) \\ 
		& PFI & 100(94$\cdot$6) & 100(95$\cdot$2) & 117(89$\cdot$9) & 127(89$\cdot$7) \\ 
		& SFI & 97(95$\cdot$1) & 83(92$\cdot$8) & 106(93$\cdot$8) & 91(94$\cdot$3) \\ 
		\hline
		\multirow{5}{*}{\textit{M4}}& Full & 100(94$\cdot$9) & 100(95$\cdot$2) & 100(94$\cdot$3) & 100(95$\cdot$2) \\ 
		& CC & 386(81$\cdot$8) & 353(86$\cdot$2) & 317(86$\cdot$4) & 308(86$\cdot$6) \\ 
		& MICE & 110(94$\cdot$8) & 114(95$\cdot$2) & 128(93$\cdot$5) & 207(86$\cdot$7) \\ 
		& PFI & 108(94$\cdot$8) & 111(94$\cdot$2) & 129(91$\cdot$6) & 197(84$\cdot$4) \\ 
		& SFI & 126(95$\cdot$0) & 124(94$\cdot$7) & 135(94$\cdot$5) & 139(94$\cdot$2) \\ 
		\hline
	\end{tabular}
\end{table}

Table \ref{tbl11} presents the relative mean squared errors and the corresponding average coverage probabilities from the above simulation study of size $B=2,000$. Under \textit{M1}, the joint model is a Gaussian mixture. The proposed method obtains almost the same RMSEs for estimating $\theta_2, \theta_3$ and $P_2$ with MICE and PFI, but outperforms MICE and PFI for estimating $P_3$. Moreover, coverage probabilities of MICE and PFI are less than 95\% in estimating $\theta_3$ and $P_3$. SFI uniformly achieves approximate 95\% coverage probabilities. Thus, we can conclude that MICE and PFI are biased, due to model misspecification. The histogram in Figure \ref{G2} shows that most of selected G values are 3, which is the true number of mixture components. 

In \textit{M2}, instead of Gaussian mixture models, one component is the exponential distribution. Table \ref{tbl11} shows that SFI outperforms MICE and PFI in term of RMSEs for estimating proportions and obtains similar performance for estimating means. The coverage rates for MICE and PFI for $P_3$ are poor.

The joint distribution \textit{M3} is a skewed distribution. From Tables \ref{tbl11}, we can see that SFI outperforms MICE and PFI uniformly. Furthermore, SFI provides better coverage probabilities than MICE and PFI  for $P_2$ and $P_3$. 

The joint distribution in \textit{M4} has a nonlinear mean structure of $Y_{i3}$. Under M4, SFI obtains the much smaller RMSE than MICE and PFI in estimating $P_4$, but larger RMSE in estimating $\theta_2, \theta_3, P_3$. However, SFI achieves consistent confidence intervals with approximate 95\% coverage probabilities. MICE and PFI are biased in interval estimation and coverage probabilities are much less than 95\%.
Overall, the performance of SFI is much better than MICE or PFI in terms of coverage probabilities.

Interestingly, the imputed estimators are sometimes more efficient than the full sample estimators. This phenomenon, called superefficiency \citep{meng1994multiple}, can happen， when the method-of-moment is used in the full sample estimator. \cite{yang2016note} give a rigorous theoretical justification for this phenomenon.

\section{Application}\label{sec:app}
\begin{figure}
	\centering
	\includegraphics[width=0.8\textwidth]{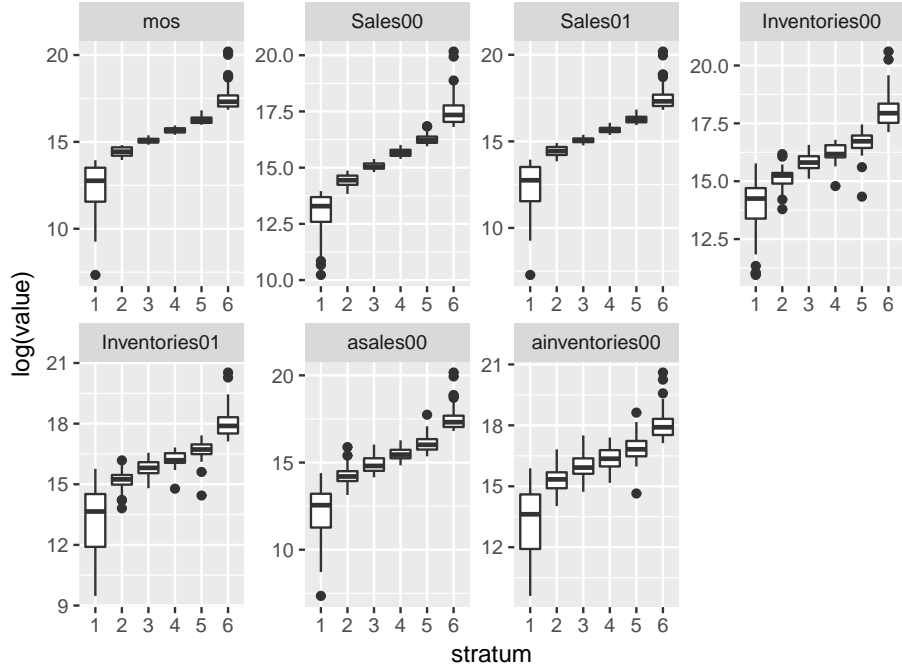}
	\captionsetup{width=0.8\textwidth}
	\caption{Overview for log-values of the monthly retail trade survey data:``mos" is frame measure of size; ``Sales00" denotes current month sales for unit (subject to missing); ``asales00" is current month administrative data value for sales; ``Sales01" means prior month sales for unit; ``Inventories00" is current month inventories for unit (Subject to missing); ``ainventories00" is current month administrative data value for inventories; ``Inventories01" is prior month inventories for unit. }\label{realdata}
\end{figure}

In this section, we apply the proposed method in \S \ref{sec:prop} to a synthetic data that mimics monthly retail trade survey data at U.S. Census Bureau. The synthetic monthly retail trade survey data was made for contest in a conference and can be found in \url{http://www.portal-stat.admin.ch/ices5/imputation-contest/}. The sampling scheme is a stratified simple random sample without replacement sample with six strata: one certain (take-all) and five non-certainty strata. The sample sizes are computed using Neyman allocation.
An overview of the monthly retail trade survey data is presented in Figure \ref{realdata}.

The overall response rate is approximately 71\%. Current month sales and inventories are subject to missingness. From Figure \ref{realdata}, we can find that this monthly retail trade data are highly skewed.  
From the normal quantile-quantile plot, normality assumption is violated and there exist three extreme outliers.

To impute current month sales and inventories, we applied the proposed fractional imputation method and MICE. After implementation, MICE failed to converge due to high correlations among the survey items.
Therefore, we only present the final analysis results using the proposed method. The final results are shown in Table \ref{result_realdata}.
\begin{table}
	{\centering
		\caption{Imputation results for the monthly retail trade survey data. Parameter estimation, 95\% confidence interval and true values are presented. 	Estimate, the estimators provided by the proposed method; Truth, the true parameter values from U.S. Census Bureau.}\label{result_realdata}
		\begin{tabular}{lrrr}
			\hline
			Parameter & Estimate  & 95\% Confidence interval &Truth \\
			\hline
			Mean of Sales00 $(\times 10^{-6})$ & 2$\cdot$28 & (2$\cdot$10, 2$\cdot$46)  &2$\cdot$30\\
			\hline
			Mean of Inventories00 $(\times 10^{-6})$ & 4$\cdot$76&  (4$\cdot$42, 5$\cdot$10)& 4$\cdot$81 \\
			\hline
			Correlation of Sales00 and Inventories00 & 0$\cdot$97& (0$\cdot$94, 0$\cdot$99)& 0$\cdot$97\\
			\hline
	\end{tabular}}
\end{table}
Comparing with the true population statistics, provided by U.S. Census Bureau, we can see that our proposed fractional imputation method with Gaussian mixture models works well to preserve the correlation structure and handle skewness and outliers.  In Table \ref{realdata}, we can see that all 95\% confidence intervals contain their true values.

\section{Discussion}\label{sec:discuss}
Fractional imputation has been proposed as a tool  for frequentist imputation, as an alternative to multiple imputation. Multiple imputation using Rubin's formula can be biased when the model is uncongenial or the point estimator is not self-efficient \citep{meng1994multiple, yang2016note}.
In this paper, we have proposed a semiparametric fractional imputation method using Gaussian mixture models to handle arbitrary multivariate missing data.  The proposed method automatically selects the size of mixture components and provides a unified framework for robust imputation.  Even if the group size $G$ increases with the sample size $n$, the resulting estimator enjoys $\sqrt{n}$-consistency. We have also extended the proposed method to incorporate categorical auxiliary variable. The flexible model assumption and efficient computation are the main advantages of our proposed method. An extension of the proposed method to survey data is a topic of future research. An R software package for the proposed method is under development.

\bibliographystyle{chicago}
\bibliography{ref}

\end{document}